\documentclass[reprint,aps,prl,superscriptaddress,amsmath,amssymb,floatfix]{revtex4-1}

\pdfoutput=1

\usepackage{graphicx}
\usepackage[pdftex,breaklinks=true,bookmarksopen=true,bookmarksopenlevel=3,bookmarksnumbered=true,colorlinks=true,urlcolor= magenta,citecolor=blue,linkcolor=blue]{hyperref}

\begin{document}

\title{Single shot phase contrast imaging using laser-produced Betatron x-ray beams}

\author{S.~Fourmaux}
\email{fourmaux@emt.inrs.ca}
\affiliation{INRS-EMT, Universit\'e du Qu\'ebec, 1650 Lionel Boulet, Varennes
J3X 1S2, Qu\'ebec, Canada}
\author{S. Corde}
\author{K. Ta Phuoc}
\affiliation{Laboratoire d'Optique Appliqu\'ee, ENSTA ParisTech -
CNRS UMR7639 - \'Ecole Polytechnique, Chemin de la Huni\`ere,
Palaiseau F-91761, France}
\author{P.~Lassonde}
\author{G.~Lebrun}
\author{S.~Payeur}
\author{F.~Martin}
\affiliation{INRS-EMT, Universit\'e du Qu\'ebec, 1650 Lionel Boulet, Varennes
J3X 1S2, Qu\'ebec, Canada}
\author{S.~Sebban}
\author{V.~Malka}
\author{A.~Rousse}
\affiliation{Laboratoire d'Optique Appliqu\'ee, ENSTA ParisTech -
CNRS UMR7639 - \'Ecole Polytechnique, Chemin de la Huni\`ere,
Palaiseau F-91761, France}
\author{J.~C.~Kieffer}
\affiliation{INRS-EMT, Universit\'e du Qu\'ebec, 1650 Lionel Boulet, Varennes
J3X 1S2, Qu\'ebec, Canada}

\begin{abstract}
Development of x-ray phase contrast imaging applications with a laboratory scale source have been limited by the long exposure time needed to obtain one image.
We demonstrate, using the Betatron x-ray radiation produced when electrons are accelerated and wiggled in the laser-wakefield cavity, that a high quality phase contrast image of a complex object (here, a bee), located in air, can be obtained with a single laser shot.
The Betatron x-ray source used in this proof of principle experiment has a source diameter of 1.7 $\mu$m and produces a synchrotron spectrum with critical energy $E_c=12.3\pm2.5\:\textrm{keV}$  and $10^{9}$ photons per shot in the whole spectrum.
\end{abstract}

\maketitle

Laser-based phase contrast x-ray imaging is a particularly attractive application of high-power laser systems for users in the biomedical field. 
This is especially true with the in-line geometry where no x-ray optics is necessary, which keeps the experiment relatively simple. 
This is realized when an object is observed in the near field Fresnel diffraction regime. 
To do so, a sufficiently small x-ray source is needed to provide x-ray radiation with high spatial coherence, and the observation detector needs to be placed at an appropriate distance behind the sample. 
The resulting image contains information on both the imaginary and the real x-ray refractive index of the object: the former corresponds to the standard x-ray radiography absorption imaging, while the latter is related to the radiation phase shift. 
Steep index transitions encountered by the x-ray beam in the object correspond to phase gradients, causing interferences and image intensity modulations that improve the object interface contrast resolution compared to standard absorption x-ray imaging. 

The in-line phase contrast imaging in the Fresnel diffraction regime has first been studied using micro-focal x-ray tubes \cite{RSI1997Pogany}.
This technique has also been demonstrated with synchrotron machines \cite{JPD2002Hwu}, $K\alpha$ laser plasma x-ray source \cite{RSI2005Toth,APL2007Laperle} and Compton x-ray sources \cite{APL2010Oliva}.  

One practical application of phase contrast imaging is small animal cancer models in vivo micro Computed Tomography ($\mu-CT$). 
The scan duration must be limited due to constraints with the animal anaesthesia to less than 10's of minutes. 
Thus, it is essential to attain a high and stable x-ray flux while keeping the x-ray source small. 
Using micro-focal x-ray tubes, the x-ray flux is reduced and it takes up to 10's of minutes to obtain one phase contrast image.
In laser-based experiments using a $K\alpha$ x-ray source, a tomographic scan would take more than one hour as the source emission is isotropic.
Synchrotron radiation sources can produce a monochromatic collimated beam of high brightness, allowing improved phase contrast imaging measurements to be made at relatively large distances but the access time to such large facilities is limited.
More recently, Compton x-ray source have demonstrated single shot phase contrast imaging but this also relies on the use of a large facility \cite{APL2010Oliva}.

High-intensity lasers can produce a femtosecond hard x-ray pulse by a more efficient process. 
Betatron x-ray beams can be generated when an intense femtosecond laser pulse, focused onto a gas jet target, interacts with the instantaneously created under-dense plasma and excites a wakefield wave in which electrons are trapped and accelerated to high energies in short distances. 
These trapped electrons incur Betatron oscillations across the propagation axis and emit x-ray photons \cite{PRL2006TaPhuoc}.

The Betatron x-ray beam is broadband, collimated within tens of milliradians with femtosecond duration. 
Its characteristics have been described in several recent experiments \cite{NatPhys2010Kneip, NJP2010Fourmaux}.
Until now,  the limited energy and brightness of this x-ray beam forced all experiments to operate under vacuum and limited imaging to simple objects, such as edges, to assess the x-ray beam properties.

In this Letter, we present what we believe to be the first demonstration of phase contrast imaging obtained in a single laser shot and in air. 
We confirm the potential of Betatron x-ray radiation for femtosecond phase contrast imaging by characterizing the x-ray source using nylon wires for calibration. 
We then demonstrate its feasibility by obtaining a phase contrast image of a bee in one single laser shot. 

In a laser-plasma accelerator, electrons are accelerated longitudinally and wiggled transversally by the electromagnetic wakefields. Experiments \cite{PRL2006TaPhuoc} have shown that Betatron oscillations occur in the wiggler regime. The integrated radiation spectrum $dI/d\omega$
is similar to that produced by synchrotrons, characterized by its critical energy $E_c$:
$dI/d\omega(x=E/E_c)\propto \int_x^\infty K_{5/3}(\xi)d\xi$, where $K_{5/3}$ is a
modified Bessel function of the second kind.

The measurements have been performed at the Advanced Laser Light Source (ALLS)
facility at INRS-EMT, using a Ti:sapphire
laser operating at 10 Hz. 

\begin{figure}[htb]
\includegraphics[width=8.5cm]{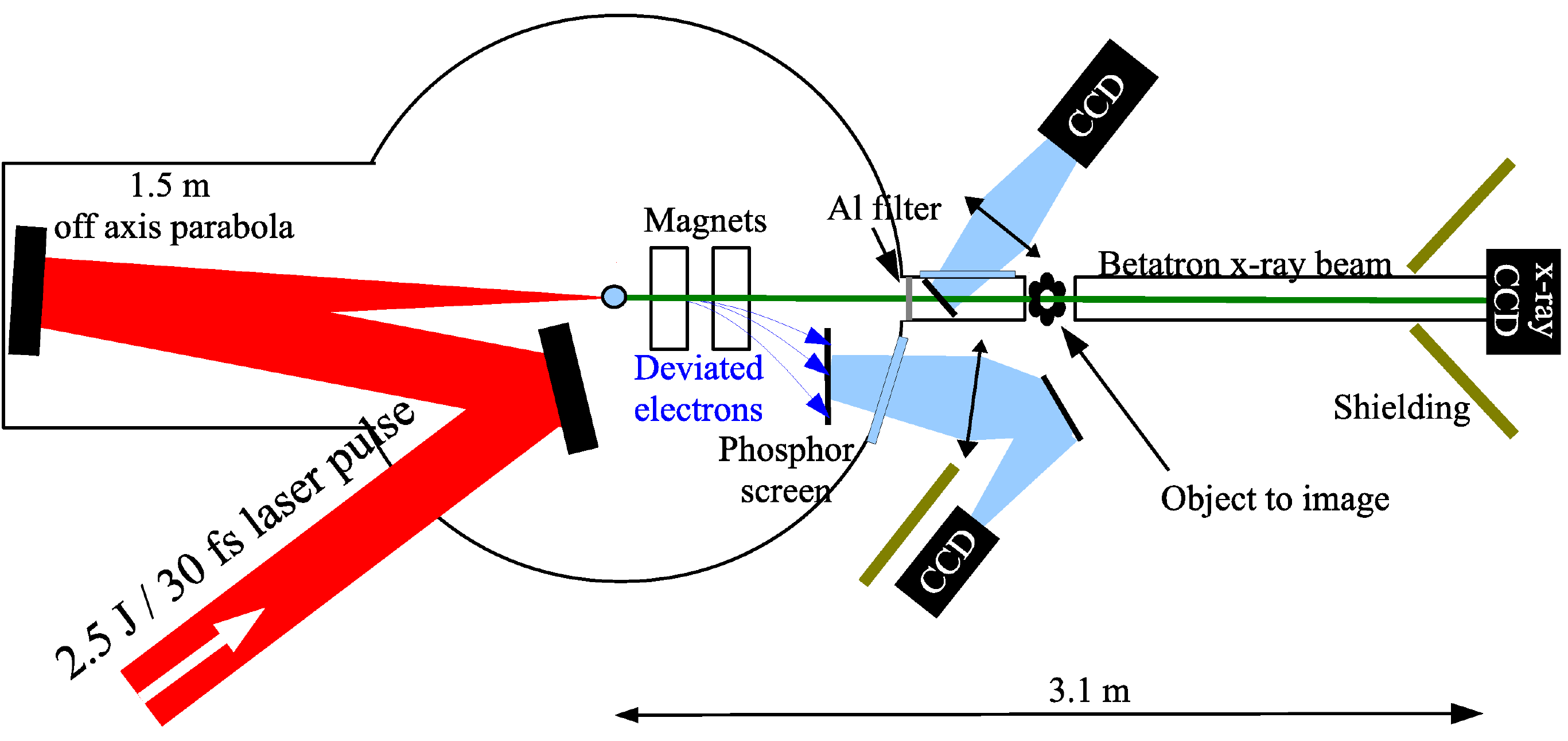}
\caption{Experimental set-up.}
\label{fig1}
\end{figure}

Figure \ref{fig1} displays the experimental configuration.
The laser system produces 2.5 J of energy on target with a full width at half maximum (FWHM) duration of 30 fs (80 TW) and linear polarization. 
The laser pulse was focused onto a supersonic helium gas jet. 
In the focal plane, the FWHM spot size was 18 $\mu\textrm{m}$ with 50\% of the total energy contained within an area limited by the $1/e^2$ radius. 
This corresponds to a laser intensity of $5\times10^{18}\:\textrm{W.cm}^{-2}$ and a normalized vector potential amplitude of $a_0=1.5$. 
For these measurements, we used a 10 mm diameter helium gas jet producing a density profile well defined by a 9-mm-long electron density plateau of $n_e=2\times10^{19}\:\textrm{cm}^{-3}$.

The electron beam produced in the interaction is measured using a permanent magnets spectrometer located inside the vacuum interaction chamber. 
Results show that electrons are accelerated up to approximately 300 MeV.
The x-rays produced by the accelerated electrons were measured by two diagnostics.
The x-ray angular profiles were measured using a GdOS:Tb phosphor screen located on the laser propagation axis (the total collection angle was 44 mrad) and recorded using a visible CCD. The measured angular spreads of the x-ray beams were $25\pm2.3$ mrad and $31\pm5$ mrad (FWHM) respectively in the horizontal and the vertical direction (averaged over 10 successive laser shots). The x-ray spectra were measured by photon counting using a deep-depletion x-ray CCD.
The details of the measurement procedure have already been described in Ref. \cite{NJP2010Fourmaux}. 
The measurements of electron spectra, x-ray angular profiles and x-ray spectra are obtained simultaneously for every single laser shot.
The phosphor used to image the x-ray profile has a hole (corresponding to 8 mrad angle) in its centre to allow the propagation of the x-rays toward the x-ray CCD for photon counting.

The x-ray CCD was placed at a distance of 3.1 m from the source and the signal was attenuated using 1.39 mm thick Al filter.
The Betatron x-ray spectrum was determined by averaging the measured signal over eight successive shots to obtain a typical spectrum. 
We obtained $2.2 \times10^8$ photons/0.1\% bandwidth/sr/shot at 10 keV.
A fit to a synchrotron distribution provides a critical energy $E_c=12.3\:\textrm{keV}$ with $\pm 2.5\:\textrm{keV}$ precision. 
The total number of photons over the whole spectrum, obtained from the synchrotron fit distribution, is $N=10^{9}$ with a confidence interval $N=8.0-13.5\times10^{8}$ (the measured solid angle is used for this calculation). 

\begin{figure}[htb]
\includegraphics[width=8.5cm]{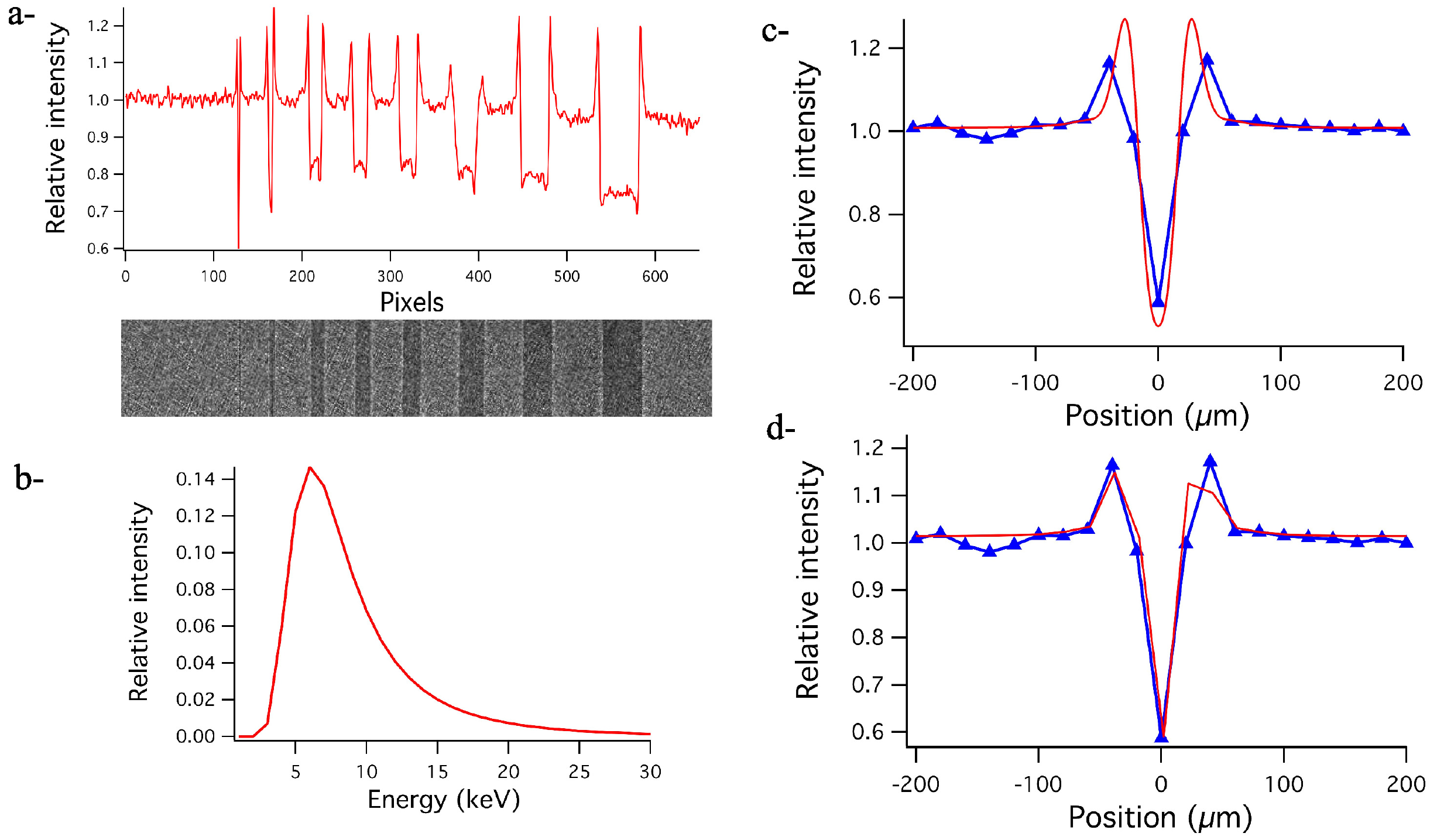}
\caption{(a) Nylon wires imaged by the Betatron x-ray beam and line out. (b) Spectrum used for the calculation. (c) and (d) Comparison of measurement (blue triangles) and calculation (red solid curve) for a 4 $\mu$m x-ray source. (d) The 20 $\mu$m CCD pixel size is taken into account.}
\label{fig2}
\end{figure} 

To assess the x-ray source potential for x-ray phase contrast imaging, we correlated the measured diffraction pattern of nylon wires of known diameter, as shown in Fig. \ref{fig2}(a). 
This is a single laser shot measurement obtained with the object located at 1.14 m from the x-ray source and the x-ray CCD used for detection. Nylon wires with diameters between 10 and 330 $\mu$m were used, all located outside the vacuum chamber (two 0.25 mm Be windows allow us to work in air).
The magnification is 2.7, the pixel size is 20 $\mu$m, which gives a measurement resolution of 7.4 $\mu$m. 
The x-ray image shows all the wires with the line-out clearly exhibiting the fringes associated with phase contrast imaging.
The Fresnel-Kirchoff integral was used in the Fresnel approximation to calculate the diffraction pattern of the 10 $\mu$m nylon wire.
The fitted synchrotron spectrum, multiplied by the spectral transmission of the absorption filters and the x-ray CCD spectral response is used in the calculation and shown in Fig. \ref{fig2}(b). In this calculation, the object absorption is considered to be thin which is correct for nylon at the energies considered here. Both measurement and calculation clearly show that using a broadband synchrotron spectrum is not a limitation to observe the first diffraction fringe. 
The calculation is in good agreement [see Figs. \ref{fig2}(c) and \ref{fig2}(d)] with our measurement when we assume an x-ray source size of 4 $\mu$m or smaller (FWHM). The comparison between the calculated diffraction pattern and the measurement is limited by the CCD spatial resolution. 

To confirm the x-ray spot size, we used a knife-edge technique.
A 1mm diameter Au microball was positioned 15 cm from the x-ray source, coupled to the x-ray CCD located at 3.1 m from the x-ray source, for a magnification of 19.7. The differentiation of the edge spread function obtained from the images yields the line spread function, which is fitted to a Gaussian distribution function. The error associated with the use of a sphere rather than an edge is estimated to be 0.6 $\mu$m, smaller than the resolution of 1 $\mu$m associated to the pixel size. This measurement yields an x-ray source size of 1.7 $\mu$m (FWHM). It is in agreement with previous experimental works \cite{PRL2006TaPhuoc,NatPhys2010Kneip,PRE2008Albert}.

The effective coherence area $A_C$ of the x-ray source can be estimated using the van Cittert-Zernike theorem. From this effective area, we can deduce the coherence length of the x-ray source as $l_c = \sqrt A_c$. Assuming 10 keV x-ray radiation, a 1.7$\mu$m source diameter and an object located at 1.1 m , we find $l_c = 9.6$  $\mu$m.
Thus, the x-ray beam is highly suitable to realize phase contrast imaging as the coherence length is similar to what can be obtain with a synchrotron machine with an object located tens of meters from the x-ray source \cite{JPD2002Hwu}.

\begin{figure}[htb]
\includegraphics[width=8.5cm]{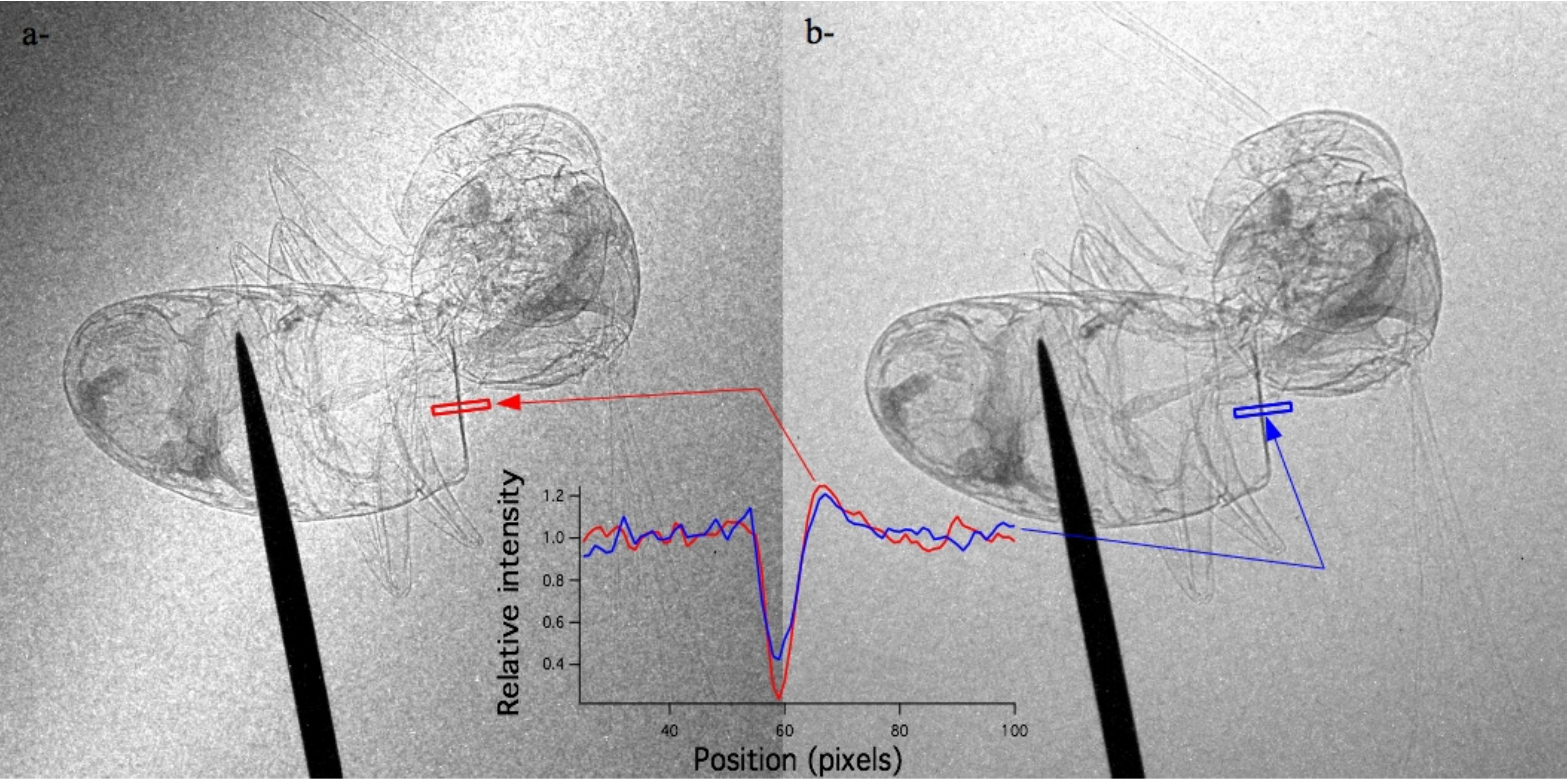}
\caption{Bee imaged with the x-ray Betatron beam with a edge line out indicated by the rectangular area: (a) 1 laser shot; (b) 13 laser shots.}
\label{fig3}
\end{figure}

Figure \ref{fig3}(a) shows a more complex and thick object: a bee located 94.5 cm from the x-ray source and imaged in air with one single Betatron x-ray pulse.
The detector used for this measurement was a GdOS:Tb phosphor with a fiber-optic faceplate coupled to a CCD camera and located 2.51 m from the x-ray source. 
Compared to direct x-ray CCD detection, the phosphor x-ray absorption optimum energy is higher: 12 keV instead of 5 keV.
The CCD is vacuum pumped and isolated from air by the faceplate. In the bee plane, the field of view is $19 \times 19$ mm$^2$ and the resolution is 15 $\mu$m.
The same bee with an accumulation of 13 pulses is shown on Fig. \ref{fig3}(b). It clearly shows a preservation of the phase contrast imaging with white and black fringes underlining some interfaces of the bee structure.

These images are comparable to measurements made using laser-based K$\alpha$ x-ray sources where thousand of shots were necessary to realize a picture \cite{RSI2005Toth}.
We can deduce the contrast $C=(H-L)/(H+L)$, where $H$ and $L$ are the signals from the edge presenting, respectively, the highest and lowest intensity. 
In single shot, the contrast value is 0.68, while with 13 shots it is reduced to 0.48, for the large fringe corresponding to the line out shown in Fig. \ref{fig3}. 
However, very thin details can be observed in single shot, but disappear in multishots because of a loss of resolution. The resolution, limited by the detector in single shot, is increased by a factor $1.4$ due to the effective multishot source size (arising from the laser pointing fluctuations).
A deconvolution from the detector resolution gives an effective source size of approximately 10-20 $\mu$m (FWHM) in multishots.

In conclusion, we have demonstrated that a phase contrast image of a bee can be recorded using  the Betatron x-ray beam produced by a single laser pulse. 
This clearly demonstrates the potential of the Betatron x-ray beam for biomedical applications, allowing tomography of small complex objects in less than a minute or imaging an object by x-ray phase contrast in real time at 10 Hz repetition rate. 
Betatron x-ray beams also allow pump and x-ray probe experiments with femtosecond time resolution which is not possible using synchrotron or Compton x-ray sources.

We thanks ALLS technical team for their support. 
The ALLS facility was funded by the Canadian Foundation for
Innovation (CFI). This work is funded by NSERC, the Canada Research Chair
program and Minist\`ere de l'\'Education du Qu\'ebec.
We acknowledge the Agence Nationale pour la Recherche, through the COKER project ANR-06-BLAN-0123-01, 
and the European Research Council through the PARIS ERC project (under Contract No. 226424) for their financial support.

\end{document}